\documentclass{ifacconf}

\usepackage{graphicx}      
\usepackage{natbib}        
\begin{document}
\begin{frontmatter}

\title{Fractional Boltzmann equation for
resonance radiation transport in plasma\thanksref{footnoteinfo}}

\thanks[footnoteinfo]{Authors are grateful to the Russian
Foundation for Basic Research (grant~10-01-00608) for financial
support.}

\author{Vladimir V. Uchaikin},
\author{Renat T. Sibatov}

\address{Ulyanovsk State University,
   Russia \\E-mail: vuchaikin@gmail.com}

\begin{abstract}
The fractional Boltzmann equation for resonance radiation
transport in plasma is proposed. We start from the standard
Boltzmann equation, averaging over frequencies leads to appearance
of fractional derivative. This fact is in accordance with the
conception of latent variables leading to hereditary and non-local
dynamics (in particular, fractional dynamics). The presence of the
fractional material derivative in the equation is concordant with
heavy tailed distribution of photon path lengths and with
spatiotemporal coupling peculiar to the process. We discuss some
methods of solution of the obtained equation and demonstrate
numerical results in some simple cases.
\end{abstract}


\end{frontmatter}


\section{Introduction}
The standard (Markovian) transport model based on the Boltzmann
equation can not describe some non-equilibrium processes called
\textit{anomalous} that take place in turbulent plasma,
interstellar magnetic fields, disordered semiconductors, and
other complex structures (see \cite{Bou:90, Met:00}). Causes of
anomality lie in non-uniformly scaled (fractal) spatial
heterogeneities, in which particle trajectories take cluster
form. Furthermore, particles can be located in some domains of
small sizes (traps) for a long time. Estimations show that path
length and waiting time distributions are often characterized
by heavy tails of the power law type. This behavior allows to
introduce time and space derivatives of fractional orders.
Distinction of path length distribution from exponential is
interpreted as a consequence of media fractality, and analogous
property of waiting time distribution as a presence of memory.

During last decades, essential progress in description of the
asymptotical behavior of such processes is achieved  by passage
from the standard diffusion equations to the equations with
fractional derivatives (see, for example, \cite{Met:00},
\cite{Sam:93}, \cite{Hil:00}, \cite{Pod:99}, \cite{Wes:02},
\cite{Nig:06}, \cite{Sib:09}, \cite{Uch:08}, \cite{Dat:10}).
The diffusion equation describes only asymptotical part of the
solution of the Boltzmann equation. For more complete
description the passage to fractional derivatives should be
performed not from the diffusion equation but on the previous
stage, i.~e. from the Boltzmann equation.

\cite{Non:89} proposed fractional generalization of the
bilinear Boltzmann equation in the form
$$
\frac{\partial f(\mathbf{r},\mathbf{v},t)}{\partial
t}=\ _0\textsf{D}^{1-\nu}_t
\left[\frac{}{}\textsf C(f,f)+S(\mathbf{r},\mathbf{v},t)\right.
$$
\begin{equation}\label{eq_Boltzmann_Nonnenm}
\left.-\mathbf{v}\frac{\partial
f(\mathbf{r},\mathbf{v},t)}{\partial
\mathbf{r}}-\frac{\mathbf{F}}{m}\frac{\partial
f(\mathbf{r},\mathbf{v},t)}{\partial \mathbf{v}}\right],
\end{equation}
where $f(\mathbf{r},\mathbf{v},t)$ is the one-particle
distribution function, $\textsf C(f,f)$ is the collision
integral, and $_0\textsf{D}^{1-\nu}_t$ is the Riemann-Liouville
partial derivative of fractional order $1-\nu$, $0<\nu<1$. The
derivation of this equation stated in \cite{Non:89} is rather
formal. It is based on the analogy between corresponding
standard diffusion equation and its fractional generalization.
Nevertheless, it leads to valid asymptotic equations whose
solutions are consistent with Monte Carlo simulation. However,
one should clearly understand what the equation describes.
Authors~\cite{Non:89} have not paid sufficient attention to
this issue. The fractional equation describes hereditary
kinetics with memory kernel of power law type which is usually
caused by waiting times in local domains with power law
asymptotes in distribution function. The process does not
assume spatial-temporal coupling: the waiting times are
considered to be independent of random free path lengths
distributed according to exponential law. Moreover, the
detailed consideration concerned only one-dimensional process.

The linear Boltzmann equation describing the anomalous
transport of neutrons within processes of scattering, fission,
and absorption, was generalized to fractional form by
\cite{Kad:10} and its solution was considered under more
realistic conditions.

We are going to consider fractional generalized kinetics in
connection with radiation transport in plasmas. The standard
Boltzmann equation has found very important applications in
studying transport of particles in plasma. It is known~(see
\cite{Per:04}, \cite{San:06}, \cite{Per:07}) that a stochastic
behavior of atomic resonance radiation migration in plasma is
of superdiffusive type. \cite{Per:04} have shown that photon
trajectories for Doppler, Lorentz, and Voight line shapes under
the assumption about complete frequency redistribution are
L\'evy flights. Superdiffusivity is a consequence of the
heavy-tailed nature of path length distribution
$$
P(\xi)=\mathrm{Prob}\{R>\xi\}\propto \xi^{-\nu}.
$$
The lighter the tail of the line shape function, the less heavy
the tail of path length pdf. \cite{San:06} concluded that this
tail is always heavy~($\nu\leq1$).

Finiteness of photon velocities assumes that the process is L\'evy
walks, not L\'evy flights. L\'evy walks give a proper dynamical
description of the superdiffusive behavior. The temporal and
spatial variables of L\'evy walks are strongly correlated.
\cite{Zol:99}, \cite{Zab:02}, \cite{Sok:03}, \cite{Chu:03},
\cite{Uch:04}, \cite{Uch:09}) proposed equations for L\'evy walks,
but these equations are obtained in diffusion approximation and
authors concentrate on the one-dimensional case.

In the present paper, we derive the fractional Boltzmann
equation for propagation of resonance radiation in plasma. We
start from the standard Boltzmann equation, which suppose
exponential distribution of path lengths of photon with a given
frequency $\omega$. Averaging over the random frequency leads
to power law distribution of path length (see \cite{Per:04})
and to appearance of fractional derivative in the equation.
This fact is in accordance with the conception of latent
variables leading to fractional dynamics~(see \cite{Uch:08_2}).
It is interesting that equation contains three-dimensional
material derivative generalizing the one-dimensional case
introduced in \cite{Sok:03}. Finally, we discuss methods of
solution of the obtained equation and present results in some
simple cases.

\section{Peculiarities of the process}

The process under consideration looks as follows. Initial
photon with frequency $\omega'$ and velocity $c$ entering the
gas goes a random free path $R$ distributed according to the
exponential law
$$
\mathrm{Prob}\{R>\xi|\omega\}=\exp(-\sigma(\omega)\xi).
$$
Then, it interacts with an atom which absorbs the photon
passing to excitement state. After a random time $T$, the atom
re-radiates a photon with a frequency $\omega$ being close to
the initial $\omega$ but not coinciding with them. The
irradiated photon continues its way as before.

Therefore, the first special feature of the process is a finite
speed of light. The second one is the random delay of the
photon re-emission. And the third, most important feature of
resonance excitation transport in a spectrum line is the fact
that frequency dependence of absorption factor $\sigma(\omega)$
repeats a shape of the very radiation spectral
line~$w(\omega)$. This leads to a specific situation: most of
irradiated photons have very short free paths and they are
absorbed again, practically at once, so that only small
fraction of photons having frequencies in tails of spectrum
(far from the center of the line) can go far from re-emission
points. We observe à L$\grave{e}$vy walk.

The resonance radiation transport theory develops in two
directions -- in light-engineering and in astrophysics. In the
first of them, the finiteness of speed of light is neglected,
that is, one accepts $c=\infty$. The main equation in this case
is the Biberman-Holstein equation. The second direction
concerns astrophysical scales, so that the traveling times are
much larger then the excitation times and the latter can be
neglected. Here we concentrate our attention on this last
problem.

\section{Integral transport equation}

For the sake of convenience, they usually identify radiated and
absorbed photons, in other words, they say that an atom
absorbing a photon radiates "the same photon" with frequency
and direction changed after an interaction with the atom, or
shorter, the photon is scattered on the atom. This process is
described by the linear Boltzmann equation
$$
\left(\frac{\partial}{\partial
t}+c\mathbf{\Omega}\nabla\right)
\Phi(\mathbf{r},\mathbf{\Omega},\omega,t)+c\sigma(\omega)
\Phi(\mathbf{r},\mathbf{\Omega},\omega,t)
$$
$$
=\int\int d\mathbf{\Omega}'d\omega'w(\omega\leftarrow\omega')
W(\mathbf{\Omega}\leftarrow\mathbf{\Omega}')c\sigma(\omega')\Phi(\mathbf{r},
\mathbf{\Omega}',\omega',t)
$$
\begin{equation}\label{eq_frac_Boltzmann_eq}
+\Phi_0(\mathbf{r},
\mathbf{\Omega},\omega)\frac{}{}\delta(t),
\end{equation}
which can be transformed to a pure integral form:
$$
\Phi(\mathbf{r},\mathbf{\Omega},\omega,t)=\Phi_0(\mathbf{r},\mathbf{\Omega},\omega,t)
$$
$$
+\int\limits_0^\infty d\xi\ e^{-\sigma(\omega) \xi}\int\int d
\mathbf{\Omega}'d\omega'\
w(\omega\leftarrow\omega')W(\mathbf{\Omega}\leftarrow\mathbf{\Omega}')
$$
$$
\times\sigma(\omega') \Phi(\mathbf{r}-\mathbf{\Omega}\
\xi,\mathbf{\Omega'},\omega',t-\xi/c).
$$
Here
$\Phi(\mathbf{r},\mathbf{\Omega},\omega,t)d\mathbf{r}d\mathbf{\Omega}d\omega$
is the mean number of photons at the moment $t$ in volume
$d\mathbf{r}=dx dy dz$ flying into solid angle
$d\mathbf{\Omega}$ with frequencies in
$[\omega,\omega+d\omega)$ Further, we use the standard
assumptions.

\begin{itemize}
    \item \textit{Assumption} 1. The complete frequency
        redistribution takes place:
$$w(\omega\leftarrow\omega')=w(\omega)$$

    \item \textit{Assumption} 2. The local thermodynamic
        equilibrium is valid:
$$
\sigma(\omega)=C w(\omega)
$$
\end{itemize}

Passing to the photon collision density
$$
F(\mathbf{r},\mathbf{\Omega},\omega,t)\equiv
c\sigma(\omega)\Phi(\mathbf{r},\mathbf{\Omega},\omega,t),
$$
we arrive at
$$
F(\mathbf{r},\mathbf{\Omega},\omega,t)=
F_0(\mathbf{r},\mathbf{\Omega},\omega,t)+
$$
$$
+\int\limits_0^\infty d\xi\ w(\omega)\sigma(\omega)
e^{-\sigma(\omega) \xi}\int d \mathbf{\Omega}'\int d\omega'\
W(\mathbf{\Omega}\leftarrow\mathbf{\Omega}')\times
$$
$$
\times F(\mathbf{r}-\mathbf{\Omega}\
\xi,\mathbf{\Omega'},\omega',t-\xi/c)
$$
Integrating over frequency
$$
F(\mathbf{r},\mathbf{\Omega},t)=\int\limits_0^\infty
F(\mathbf{r},\mathbf{\Omega},\omega,t) d\omega,
$$
we obtain the following integral equation
$$
F(\mathbf{r},\mathbf{\Omega},t)-F_0(\mathbf{r},\mathbf{\Omega},t)=
$$
$$
=\int\limits_0^\infty d\xi\ p(\xi)\int d \mathbf{\Omega}'
W(\mathbf{\Omega}\leftarrow\mathbf{\Omega}')\
F(\mathbf{r}-\mathbf{\Omega}\ \xi,\mathbf{\Omega'},t-\xi/c),
$$
where the kernel is given by
$$
p(\xi)=\left\langle\sigma(\omega) e^{-\sigma(\omega)
\xi}\right\rangle=\int\limits_0^\infty w(\omega) \sigma(\omega)
e^{-\sigma(\omega) \xi} d\omega.
$$

Characteristic feature of resonance radiation propagation is the
fact that absorption spectrum (as an approximation of
thermodynamic equilibrium and total redistribution of frequency)
is proportional to the spectrum of radiation
$$
\sigma(\omega)=C w(\omega)
$$
Calculations lead us to the following asymptotic properties of
the path length distribution
$$
P(\xi)=\left\langle e^{-\sigma(\omega) \xi}\right\rangle \sim
\frac{\xi^{-\nu}}{\gamma\Gamma(1-\nu)},
$$
$$
p(\xi)=-P'(\xi)=\left\langle \sigma e^{-\sigma(\omega)
\xi}\right\rangle \sim \frac{
\xi^{-\nu-1}}{\gamma\Gamma(1-\nu)}.
$$
To satisfy this condition, we choose the distribution in the
form of "fractional exponents":
\begin{equation}\label{eq_frac_exp}
p(\xi)=\gamma \xi^{\nu-1} E_{\nu,\nu}(-\gamma \xi^\nu),
\end{equation}
where $E_{\alpha,\beta}(x)$ is the two-parameter Mittag-Leffler
function (see, for example \cite{Uch:08}).

\section{Fractional Boltzmann equation}

Generalized Boltzmann equation for the case of path length
distribution (\ref{eq_frac_exp}) is obtained in the form
$$
\left(\frac{\partial}{\partial
t}+c\mathbf{\Omega}\nabla\right)^\nu
\Phi(\mathbf{r},\mathbf{\Omega},t)+\mu
\Phi(\mathbf{r},\mathbf{\Omega},t)=
$$
$$
=\mu\int
W(\mathbf{\Omega}\leftarrow\mathbf{\Omega}')\Phi(\mathbf{r},
\mathbf{\Omega}',t)d\mathbf{\Omega}'+
$$
\begin{equation}\label{eq_frac_Boltzmann_eq}
+\left(\frac{\partial}{\partial
t}+c\mathbf{\Omega}\nabla\right)^{\nu-1}\left[\Phi_0(\mathbf{r},
\mathbf{\Omega})\frac{}{}\delta(t)\right],
\end{equation}
where $\mu=c\gamma$ and the operator
$$
\left(\frac{\partial}{\partial
t}+c\mathbf{\Omega}\nabla\right)^\nu
\Phi(\mathbf{r},\mathbf{\Omega},t)=
$$
\begin{equation}\label{eq_frac_mat_der}
=\frac{1}{\Gamma(1-\nu)}\left(\frac{\partial}{\partial
t}+c\mathbf{\Omega}\nabla\right)\int\limits_0^t
\frac{\Phi(\mathbf{r}-c\mathbf{\Omega}(t-\tau),\mathbf{\Omega},\tau)}{(t-\tau)^\nu}d\tau
\end{equation}
is the fractional generalization of the material derivative.

Ordinary material derivative is determined as
$$
\left(\frac{\partial}{\partial
t}+\mathbf{c}\frac{\partial}{\partial
\mathbf{r}}\right)f(\mathbf{r},t)=\lim\limits_{h\downarrow0}
\frac{f(\mathbf{r},t)-f(\mathbf{r}-\mathbf{c}h,t-h)}{h}.
$$
The operator
$$
A=-\left(\frac{\partial}{\partial
t}+\mathbf{c}\frac{\partial}{\partial \mathbf{r}}\right)
$$
is the infinitesimal operator generating the semigroup of
operators
$$
T_h f(\mathbf{r},t)=f(\mathbf{r}-\mathbf{c}h,t-h)
$$
Consequently, we have for $(-A)^\alpha$
$$
(-A)^\alpha f(\mathbf{r},t)=\left(\frac{\partial}{\partial
t}+\mathbf{c}\frac{\partial}{\partial \mathbf{r}}\right)^\alpha
f(\mathbf{r},t)=
$$
$$
=\lim\limits_{h\downarrow0}\
\frac{\alpha}{\Gamma(1-\alpha)}\int\limits_h^\infty
s^{-\alpha-1}\left[f(\mathbf{r},t)-f(\mathbf{r}-\mathbf{c}s,t-s)\frac{}{}\right]ds
$$

By integrating by parts, the last operator can be reduced to the
form
$$
\left(\frac{\partial}{\partial
t}+\mathbf{c}\frac{\partial}{\partial \mathbf{r}}\right)^\alpha
f(\mathbf{r},t)=
$$
$$
=\frac{1}{\Gamma(1-\alpha)}\left(\frac{\partial}{\partial
t}+\mathbf{c}\frac{\partial}{\partial
\mathbf{r}}\right)\int\limits_{-\infty}^t
\frac{f(\mathbf{r}-\mathbf{c}(t-\tau),\tau)}{(t-\tau)^\alpha}d\tau
$$

If we consider the case of a space-homogeneous distribution of
particles
$\Phi(\mathbf{r},\mathbf{\Omega},t)\equiv\Phi(\mathbf{\Omega},t)$
this operator becomes the fractional Rieman-Liouville
derivative of order $\nu$
$$
\left(\frac{\partial}{\partial
t}+c\mathbf{\Omega}\nabla\right)^\nu
\Phi(\mathbf{r},\mathbf{\Omega},t) \mapsto\ _0\textsf{D}^\nu_t
\Phi(\mathbf{\Omega},t).
$$

In a stationary problem, when the function
$\Phi(\mathbf{r},\mathbf{\Omega})$ does not depend on time,
$$
\left(\frac{\partial}{\partial
t}+c\mathbf{\Omega}\nabla\right)^\nu
\Phi(\mathbf{r},\mathbf{\Omega}) \mapsto\ c (\mathbf{\Omega}
\nabla)^\nu\Phi(\mathbf{r},\mathbf{\Omega},t),
$$
this operator represents fractional generalization of the
directional derivative.

\section{Solutions of fractional Boltzmann equation}

\subsection{Analytical results}

In one-dimensional case, when photons can fly only along of
$x$-axis, the fractional generalization of the Boltzmann
equation is of the form
$$
\left(\frac{\partial}{\partial t}+c \Omega_x\frac{\partial}{\partial
x}\right)^\nu \Phi(x,\Omega_x,t)+\mu \Phi(x,\Omega_x,t)=
$$
$$
=\mu J(x,\Omega_x,t)+\left(\frac{\partial}{\partial t}+c
\Omega_x\frac{\partial}{\partial x}\right)^{\nu-1}S(x,\Omega_x,t).
$$
Using expressions for the collision integral
$$
J(x,\Omega_x,t)=\frac{1}{2}\left[\Phi_+(x,t)+\Phi_-(x,t)\right]\
[\delta(\Omega_x-1)+\delta(\Omega_x+1)]
$$
and for the one-dimensionally "isotropic" instantaneous source
$$
S(x,\Omega_x,t)=\frac{1}{2}\
[\delta(\Omega_x-1)+\delta(\Omega_x+1)]\delta(x)\delta(t)
$$
we obtain the following equations:
$$
\left(\frac{\partial}{\partial t}+c\frac{\partial}{\partial
x}\right)^\nu \Phi_+(x,t)=
$$
$$
=\frac{\mu}{2}[\Phi_-(x,t)-\Phi_+(x,t)]+\frac{t^{-\nu}}{2\Gamma(1-\nu)}\
\delta(x-ct)
$$
$$
\left(\frac{\partial}{\partial t}-c\frac{\partial}{\partial
x}\right)^\nu
\Phi_-(x,t)=
$$
$$
=\frac{\mu}{2}[\Phi_+(x,t)-\Phi_-(x,t)]+\frac{t^{-\nu}}{2\Gamma(1-\nu)}\
\delta(x+ct)
$$
$$
\Phi(x,t)=\Phi_+(x,t)+\Phi_-(x,t)
$$
From last equations one follows{\small
$$
\left[\left(\frac{\partial}{\partial t}-c\frac{\partial}{\partial
x}\right)^\nu+\left(\frac{\partial}{\partial
t}+c\frac{\partial}{\partial x}\right)^\nu
+\frac{2}{\mu}\left(\frac{\partial^2}{\partial
t^2}-c^2\frac{\partial^2}{\partial
x^2}\right)^\nu\right]\Phi(x,t)=
$$}
$$
=\frac{t^{-\nu}}{2\Gamma(1-\nu)}\
[\delta(x-ct)+\delta(x+ct)]-
$$
\begin{equation}\label{eq_telegraph}
-\frac{\nu\ t}{\mu\ [\Gamma(1-\nu)]^2}\left(\frac{c^2 t^2 -x^2}{4
c^2}\right)^{-1-\nu}.
\end{equation}

In the asymptotics of large times the last equation passes into
the following one
$$
\left[\left(\frac{\partial}{\partial t}-c\frac{\partial}{\partial
x}\right)^\nu+\left(\frac{\partial}{\partial
t}+c\frac{\partial}{\partial x}\right)^\nu \right]\Phi(x,t)
=
$$
$$
=\frac{t^{-\nu}}{2\Gamma(1-\nu)}\ [\delta(x-ct)+\delta(x+ct)],
$$
solution of which is expressed in terms of elementary functions
(see \cite{Uch:09})
$$
\Phi(x,t)=\frac{2 \sin\pi\nu}{\pi}\times
$$
$$
\times\frac{\left(1-x^2/c^2
t^2\right)^{\nu-1}}{(1-x/ct)^{2\nu}+(1+x/ct)^{2\nu}+2\left(1-x^2/c^2
t^2\right)^\nu \cos\pi\nu}.
$$

Quantitative analysis of distributions can be performed using the
method of moments. Due to finiteness of photon velocity, its
position is bounded random variable and according to Cramer's
criterion distribution of excitations is unambiguously determined
by the moments of this distribution. Moments can be found from a
characteristic function.

For the one-dimensional case, the Fourier-Laplace transformation
of the equation~(\ref{eq_telegraph}) leads to the following
expression
$$
\widetilde{\Phi}(k,\lambda)=\frac{(\lambda-ick)^{\nu-1}+(\lambda+ick)^{\nu-1}
+\displaystyle\frac{2\lambda}{\mu}\left(\lambda^2+c^2
k^2\right)^\nu}{(\lambda-ick)^{\nu}+(\lambda+ick)^{\nu}+\displaystyle
\frac{2\lambda}{\mu}\left(\lambda^2+c^2
k^2\right)^\nu}.
$$
In the asymptotics of small $k$ and $\lambda$ corresponding to the
diffusion limit, we obtain
$$
\widetilde{\Phi}(k,\lambda)\sim\frac{(\lambda-ick)^{\nu-1}+(\lambda+ick)^
{\nu-1}}{(\lambda-ick)^{\nu}+(\lambda+ick)^{\nu}}.
$$
Moments are determined by the relation
$$
\tilde{m}_n(\lambda)=(-i)^n \frac{\partial^n
\Phi(k,\lambda)}{\partial k^n}.
$$

In isotropic case, Fourier-Laplace transformation of
(\ref{eq_frac_Boltzmann_eq}) leads to the following asymptotical
form for $\widetilde{\Phi}(\mathbf{k},t)$
$$
\widetilde{\Phi}(\mathbf{k},t)\sim \frac{\int \left(\lambda-ic
\mathbf{\Omega}\hspace{0.5mm}
\mathbf{k}\right)^{\nu-1}d\mathbf{\Omega}}{\int \left(\lambda-ic
\mathbf{\Omega}\hspace{0.5mm}
\mathbf{k}\right)^{\nu}d\mathbf{\Omega} }.
$$

\subsection{Numerical results}

Here, a computational algorithm for the process is described,
realized in the Monte Carlo code and used for computing the
resonance radiation and excitation distribution in plasma. An
important advantage of the method is simplicity of taking into
account boundary conditions and inhomogeneity of the medium.

This method represents randomized scheme of realization of
generation expansion. It can be considered as statistical
simulation of transport process characterized by known
distributions of independent random parameters determining a
random trajectory. Set of large number of independent trajectories
allows to estimate statistically a process quality interesting for
us.

\begin{figure}[tbh]
\centering
\includegraphics[width=0.45\textwidth]{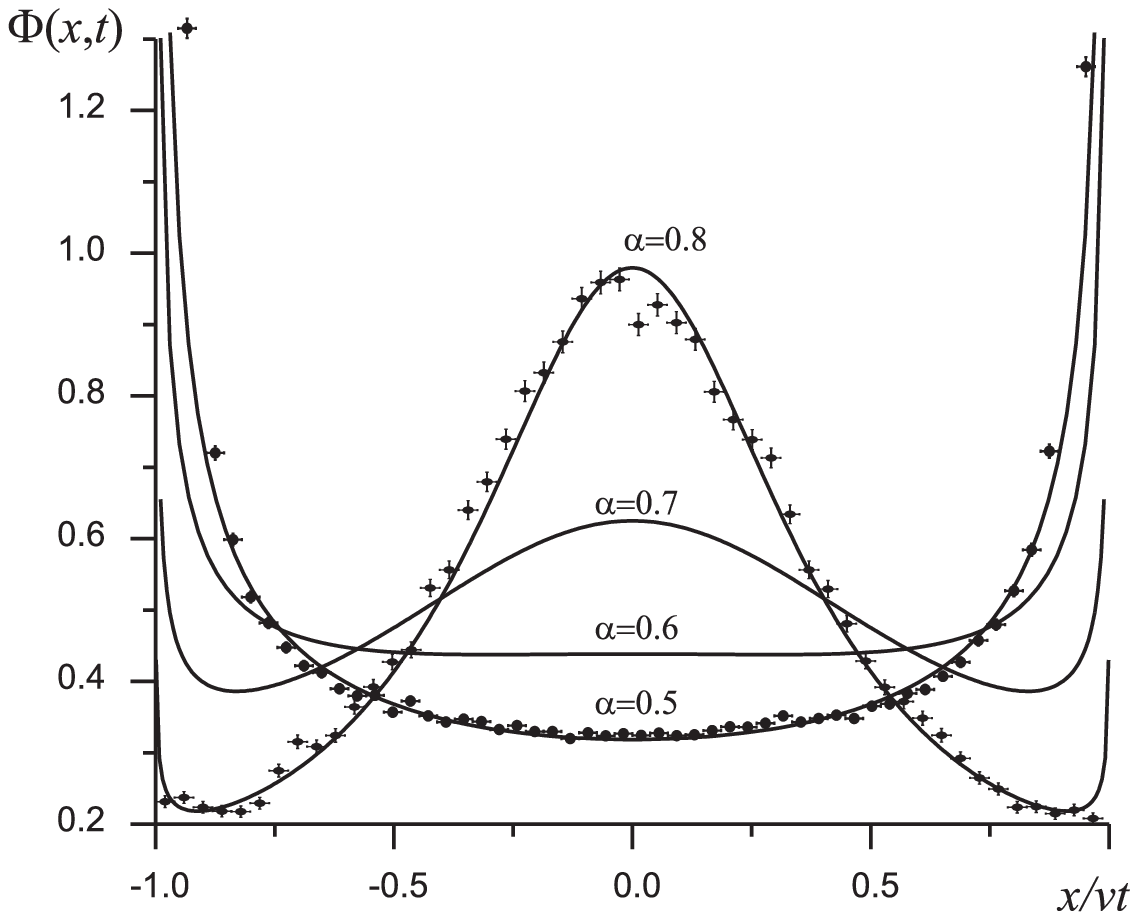}\hspace{1cm}
\caption{Comparison of analytical solutions with the results of
Monte Carlo simulation for isotropically one-dimensional
case as functions of dimensionless variable $x/ct$.}\label{fig_1D}
\end{figure}

\begin{figure}[tbh]
\centering
\includegraphics[width=0.4\textwidth]{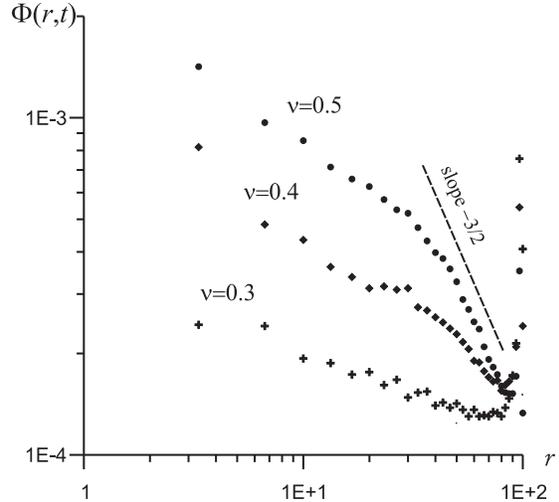}\hspace{1cm}
\caption{Results of Monte Carlo simulation for isotropic 3D
case.}\label{fig_3D}
\end{figure}

There is the typical sequence of operations for simulation of
photon trajectory.

\begin{description}
    \item[--] A position of excited atom at the initial
    time moment is chosen. PDF of its radius-vector
    $\mathbf{R}_0$ is determined by given initial distribution of
    excitations $N(\mathbf{r},0)$ through the relation ${p_{\mathbf{R}}=N(\mathbf{r},0)/\int
    N(\mathbf{r},0)d\mathbf{r}}$.

    \item[--] The random time till the moment of photon emission by the atom
    is generated according to the exponential distribution $p_T(t)=\gamma e^{-\gamma
    t}$.

    \item[--] The frequency of emitted photon is distributed with the
    pdf $w(\omega)$.

    \item[--] A direction $\mathbf{\Omega}$ of photon
    emission is generated. This direction is characterized by random variables
    $\cos \Theta$ and $\Psi$, which are uniformly distributed (in isotropic case)
    in $[-1,1]$ and
    $[0,2\pi)$, respectively.
    \item[--] A photon path length $R$ is simulated in infinite
    homogeneous medium, $p_R(\xi)=\sigma(\omega)e^{-\sigma(\omega)\xi}$. If the whole
    segment $[\mathbf{R}_0,\mathbf{R}_0+\mathbf{\Omega}R]$ keeps
    within the volume occupied by plasma, its second end is taken
    as the position of another excited atom. This atom absorbs the photon.
    If this segment traverses absorbing or transparent boundary, the trajectory
    comes to the end.
    \item[--] For new excited atom, waiting time till the moment of photon emission
    is simulated.
    And so on, described operations are
    repeated before the end of trajectory.
\end{description}

From physical point of view, the fractional operator in the
Boltzmann equation means that anomalously long free paths arising
from time to time tear the trajectory into more or less localised
clusters which look separated at any scale.

Solutions obtained by Monte Carlo simulations are presented in
Fig.~\ref{fig_1D},~\ref{fig_3D} (points). One-dimensional
solutions listed in the previous subsection agree with the
numerical results.

\section{Conclusion}

The fractional generalization of the Boltzmann equation taking
into account spatiotemporal coupling peculiar to resonance
radiation transport is proposed. This coupling is provided by
heavy tailed distribution of photon path lengths which arises
after averaging over frequencies and by finiteness of photon
velocity. Heavy tails of path length distribution leads to
appearance of fractional derivative in the kinetic equation, but
due to coupling, this fractional operator represents not simple
fractional Laplacian. It is a fractional analogue of the material
derivative that is agree with results obtained by \cite{Sok:03}
for one-dimensional L\'evy walks.

\end{document}